\documentclass[aps,prl,reprint,nofootinbib,showpacs,superscriptaddress,twocolumn,longbibliography]{revtex4-1}

\usepackage{amssymb,amsmath,amsthm, amsfonts,mathtools,graphicx,xcolor,braket,algorithm,algpseudocode,latexsym}
\usepackage[caption=false]{subfig}
\usepackage{hyperref}
\hypersetup{colorlinks=true,allcolors=blue}
\newtheorem{theorem}{Theorem}
\newtheorem{conj}[theorem]{Conjecture}
\newtheorem{lemma}[theorem]{Lemma}
\newcommand{\norm}[1]{\left\lVert#1\right\rVert}
\makeatletter
\newcommand{\algmargin}{\the\ALG@thistlm}   
\makeatother
\algnewcommand{\parState}[1]{\State%
    \parbox[t]{\dimexpr\linewidth-\algmargin}{\strut #1\strut}}
\usepackage{indentfirst}
\begin{document}

\title{Learning Unitaries by Gradient Descent}

\author{Bobak Toussi Kiani}
\affiliation{Massachusetts Institute of Technology, 77 Massachusetts Avenue, Cambridge, MA 02139, USA}
\author{Seth Lloyd}
\affiliation{Massachusetts Institute of Technology, 77 Massachusetts Avenue, Cambridge, MA 02139, USA}
\author{Reevu Maity}
\affiliation{Clarendon Laboratory, University of Oxford, Parks Road, Oxford, OX1 3PU, UK}

\begin{abstract}

We study the hardness of learning unitary transformations in $U(d)$ via gradient descent on time parameters of alternating operator sequences. We provide numerical evidence that, despite the non-convex nature of the loss landscape, gradient descent always converges to the target unitary when the sequence contains $d^2$ or more parameters. Rates of convergence indicate a ``computational phase transition." With less than $d^2$ parameters, gradient descent converges to a sub-optimal solution, whereas with more than $d^2$ parameters, gradient descent converges exponentially to an optimal solution.

\end{abstract}

\maketitle

\bigskip

\section{Introduction}

A fundamental task in both quantum computation and quantum control is to determine the minimum amount of resources required to implement a desired unitary transformation. In this paper, we present a simple model that allows us to analyze key aspects of implementing unitaries in the context of both quantum circuits and quantum control. In particular, we implement unitaries using sequences of alternating operators of the form $e^{-i At_K} e^{-i B \tau _K}  \ldots e^{-i At_1} e^{-i B \tau _1}$. Each unitary is parameterized by the times $\{t_1, \tau_1, \ldots, t_K, \tau_K\}$. This approach of parameterizing unitaries is the basis for the quantum approximate optimization algorithm (QAOA) \cite{farhi2014quantum1,farhi2014quantum2}. The acronym QAOA is also used to refer to the phrase ``Quantum Alternating Operator Ansatz." Recently, it has been shown that quantum alternating operator unitaries can perform universal quantum computation \cite{lloyd2018quantum}. In the infinitesimal time setting, QAOA also encompasses the more general problem of the application of time varying quantum controls \cite{rabitz2004quantum,khaneja2005optimal,rabitz2005landscape,rabitz2006topology,chakrabarti2007quantum,moore2008relationship,rabitz2009landscape,brif2010control,riviello2015searching,riviello2017searching,russell2016quantum}. In this work, we study the quantum alternating operator formalism as a general framework of performing arbitrary unitary transformations.  

We investigate the difficulty of learning Haar random unitaries in $U(d)$ using parameterized alternating operator sequences. Here, we find that unsurprisingly, when the number of parameters in the sequence is less than $d^2$, gradient descent fails to learn the random unitary. 
Initially, we had expected that because of the highly non-convex nature of the loss landscape, when the number of parameters in the sequence was greater than or equal to $d^2$ -- the minimum number of parameters required to specify a $d \times d$ unitary matrix -- gradient descent would sometimes fail to learn the target unitary. However, our numerical experiments reveal the opposite. When the number of parameters is $d^2$ or greater, gradient descent \textit{always} finds the target unitary. Moreover, we provide evidence for a ``computational phase transition" at the critical point between the under-parameterized and over-parameterized cases where the number of parameters in the sequence equals $d^2$. 

\emph{Learning Setting.} Suppose we have knowledge of the entries of a unitary $\mathcal{U} \in U(d)$ and access to the Hamiltonians $\pm A$ and $\pm B$. Recent work has provided a constructive approach to build a learning sequence $\mathcal{V}(\vec{t},\vec{\tau}) = e^{-i At_K} e^{-i B \tau _K}  \ldots e^{-i At_1} e^{-i B \tau _1}$ that can perform any target unitary $\mathcal{U}$ where $K = O(d^2)$  \cite{lloyd2019efficient}. In this work, we ask whether  optimal learning sequences for performing the target unitary $\mathcal{U}$ can be obtained by using gradient descent optimization on the parameters $\vec{t}, \vec{\tau}$ of $\mathcal{V}(\vec{t}, \vec{\tau})$. The matrices $A,B$ are sampled from the Gaussian Unitary Ensemble (GUE) so that the algebra generated by $A,B$ via commutation is  with probability one complete in $u(d)$, \textit{i.e.}, the system is controllable \cite{rabitz2004quantum,khaneja2005optimal,rabitz2005landscape,rabitz2006topology,chakrabarti2007quantum,moore2008relationship,rabitz2009landscape,brif2010control,riviello2015searching,riviello2017searching,russell2016quantum}. The parameters $\vec{t},\vec{\tau}$ represent the times for which the generators of $\mathcal{V}(\vec{t},\vec{\tau})$ are applied. We assume we can apply  $\pm A, \pm B$; equivalently, we can take $t_j, \tau_j$ to be positive or negative. Note that this problem formulation lies in the domain of quantum optimization algorithms such as the Quantum Approximate Optimization Algorithm \cite{farhi2016quantum,jiang2017qaoa,zhou2018quantum,gilyen2019optimizing}, the Variational Quantum Eigensolver \cite{mcclean2016theory,peruzzo2014variational,khatri2019quantumassisted,sharma2019noise}, and the Variational Quantum Unsampling \cite{carolan2020variational} in which one varies the classical parameters in a quantum circuit to minimize some objective function.

In general, the control landscape for learning the unitary $\mathcal{U}$ is highly non-convex \cite{rabitz2004quantum,khaneja2005optimal,rabitz2005landscape,rabitz2006topology,chakrabarti2007quantum,moore2008relationship,rabitz2009landscape,brif2010control,riviello2015searching,riviello2017searching,russell2016quantum}. Gradient descent algorithms do not necessarily converge to a globally optimal solution in the parameters of a non-convex space \cite{zaheer2018adaptive}, and they frequently converge instead to some undesired critical point of the loss function landscape. 
We study how hard it is to learn an arbitrary unitary with the quantum alternating operator formalism via gradient descent. We quantify the hardness of learning a unitary with the minimum number of parameters required in the sequence $\mathcal{V}(\vec{t},\vec{\tau})$ to perform the unitary $\mathcal{U}$. Since $\mathcal{U}$ has $d^2$ independent parameters, in general, at least $d^2$ parameters in the sequence $\mathcal{V}(\vec{t},\vec{\tau})$ are required to learn a unitary $\mathcal{U} \in  U(d)$ within a desired error. Nevertheless, the non-convex loss landscape suggests that it might not be possible to learn an arbitrary $\mathcal{U}$ with gradient descent using $O(d^2)$ parameters. Our work numerically shows that exactly $d^2$ parameters in the sequence $\mathcal{V}(\vec{t},\vec{\tau})$ suffice to learn an arbitrary unitary $\mathcal{U}$ to a desired accuracy. 

We also consider the case of learning ``shallow" target unitaries of the form $\mathcal{U}(\vec{t},\vec{\tau}) = e^{-i At_N} e^{-i B \tau _N}  \ldots e^{-i At_1} e^{-i B \tau _1}$ where the number of parameters in the target unitary is $2N \ll d^2$. For example, the simplest such target unitary is a depth-1 sequence $\mathcal{U}(t,\tau) = e^{-i A t} e^{-i B \tau}$. Such unitaries are, by definition, attainable via a shallow depth alternating operator sequence, and we look to see if it is possible to use gradient descent to obtain a learning sequence $\mathcal{V}(\vec{t},\vec{\tau})$ of the same depth that approximates the target unitary $\mathcal{U}(\vec{t},\vec{\tau})$. That is, we look at the alternating operator version of whether it is possible to learn the unitaries generated by shallow quantum circuits. We find that gradient descent typically requires $d^2$ parameters in the sequence $\mathcal{V}(\vec{t},\vec{\tau})$ to learn even a depth-1 unitary. This result suggests that gradient descent is not an efficient method to learn low depth unitaries.


Rabitz \textit{et al.} consider the case of controllable quantum systems with time varying controls, including systems with drift, and show that when the controls are unconstrained (space of controls is essentially infinite dimensional), there are no sub-optimal local minima even though loss landscapes may be non-convex \cite{rabitz2004quantum, rabitz2005landscape, russell2016quantum}. For example, it has been shown in \cite{rabitz2009landscape} that non-convexity in the loss landscape of fully controllable quantum systems with infinite dimensional control fields is due to the presence of non-trapping saddle points in the loss landscape. When the sequence of controls is finite dimensional, prior studies sometimes find traps in the control landscape \cite{moore2008relationship,riviello2015searching, riviello2017searching}. Here, we look at the simplest possible case where the system does not have drift and the space of controls is finite dimensional. Our numerical results show that even in spaces where the dimension of the system is the minimum it can be to attain the desired unitary and the control landscape is highly non-convex, it still contains no sub-optimal local minima and gradient descent obtains the global optimal solution.


We now provide a detailed numerical analysis of the learnability of both arbitrary and shallow depth unitaries using gradient descent optimization.

   \section{Numerical experiments for learning an arbitrary unitary}
   
   \begin{figure}[t]
   \includegraphics[width=1\linewidth]{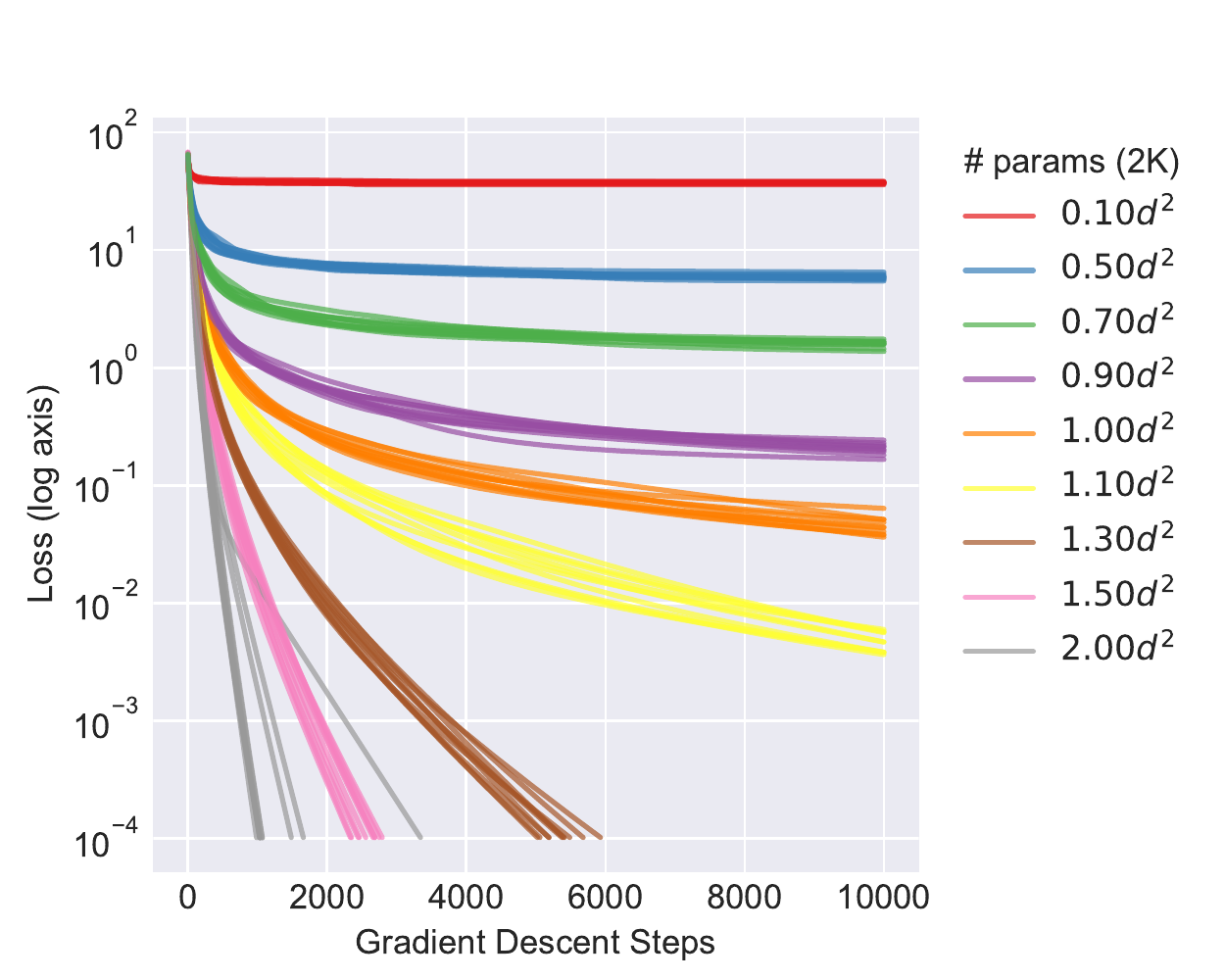}%
   \caption{Gradient descent experiments for a Haar random target unitary $\mathcal{U}$ of dimension 32. The logarithm of the loss function $L(\vec{t},\vec{\tau})$ with increasing gradient descent  steps for learning sequences $\mathcal{V}(\vec{t},\vec{\tau})$ with $2K$ parameters.}
  \label{fig1}
   \end{figure}
   
   \begin{figure}[!ht]
   
       \subfloat[]{%
       \includegraphics[width=1\linewidth]{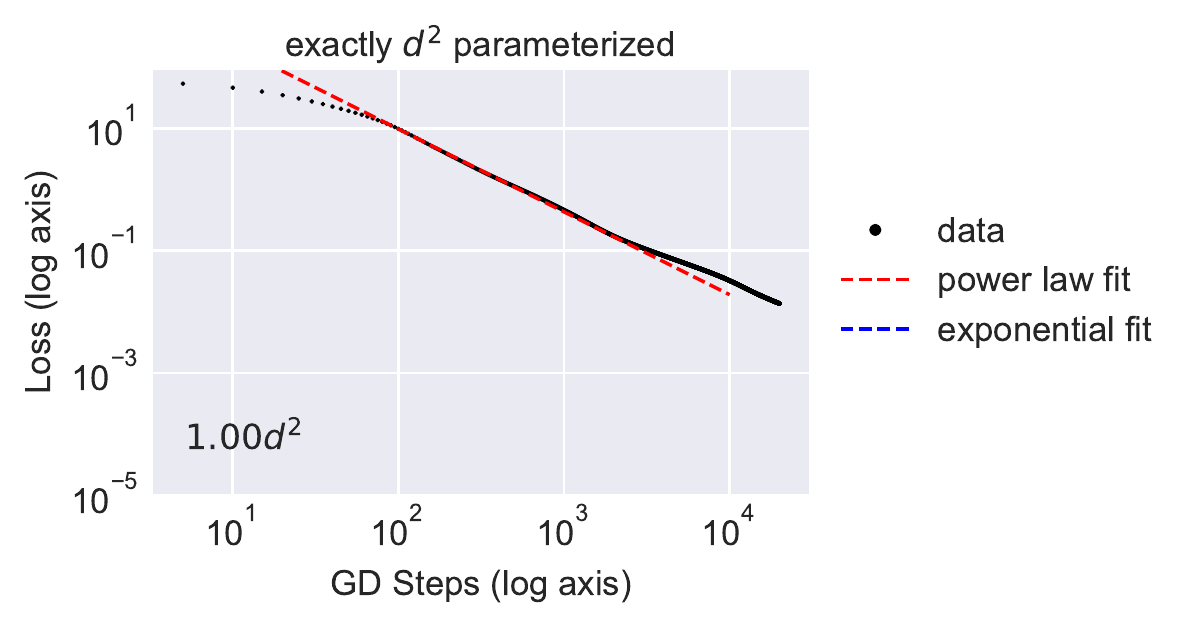}%
       }
       
       \subfloat[]{%
       \includegraphics[width=1\linewidth]{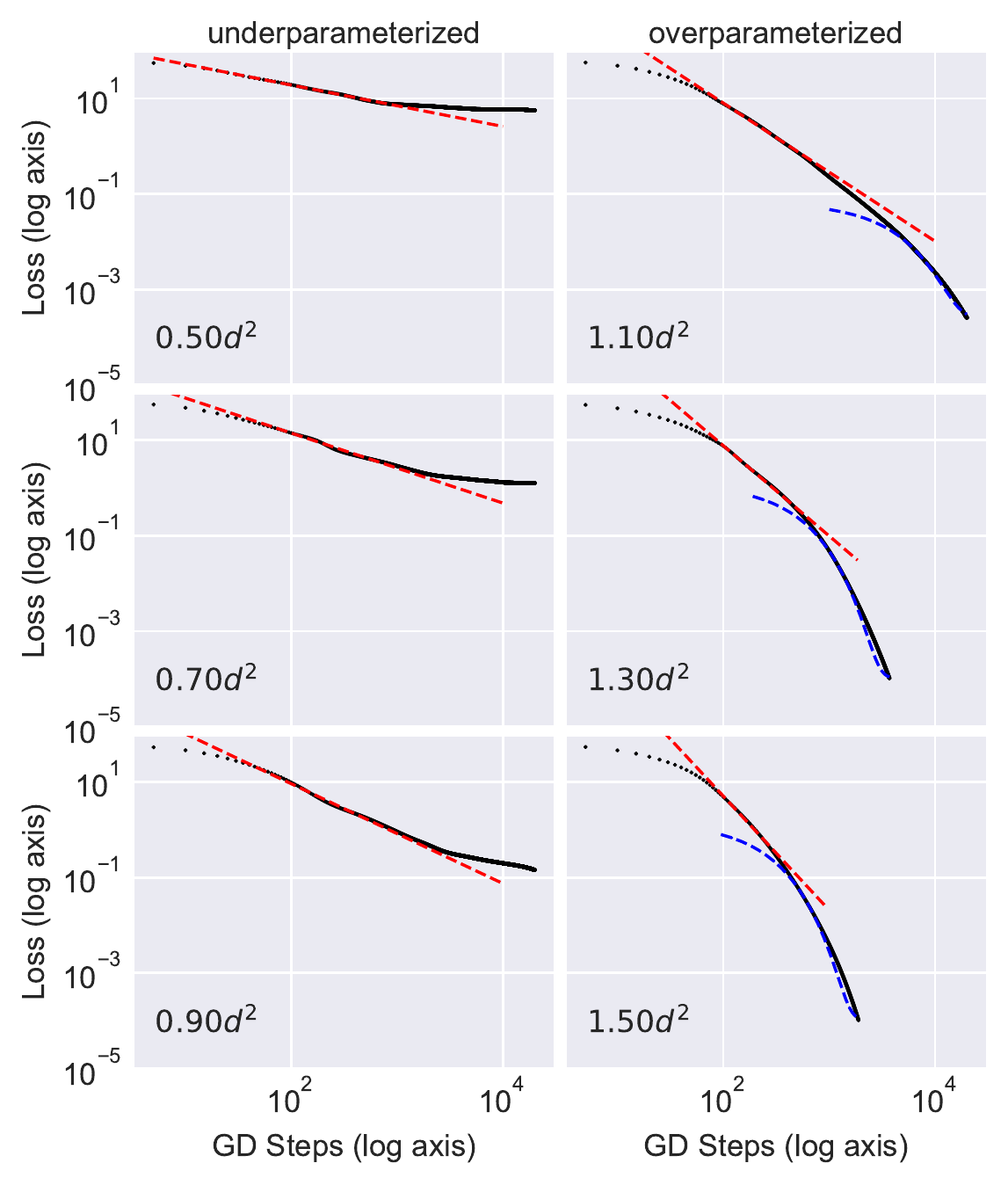}%
       }
       
   \caption{Gradient descent experiments for a Haar random target unitary $\mathcal{U}$ of dimension 32 exhibit a power law convergence in the first 1,000 gradient descent steps (best fit line shown in dashed red in the plot). In the under-parameterized case, at a certain point, the gradient descent plateaus at a sub-optimal local minimum. In the over-parameterized case, after the power law regime, the gradient descent enters an exponential regime consistent with a quadratic form for the loss function in the vicinity of the global minimum (best fit line shown in dashed blue in the plot). In the critical case, $2K = d^2$, the power law persists throughout the gradient descent providing further evidence for a computational phase transition.} 
   \label{fig2} 
   \end{figure}

   In this section, we present numerical experiments that aim to learn an arbitrary unitary $\mathcal{U}$ by constructing a sequence $\mathcal{V}(\Vec{t}, \Vec{\tau}) = e^{-i At_K} e^{-i B \tau _K}  \ldots e^{-i At_1} e^{-i B \tau _1} $ and performing gradient descent on all $2 K$ parameters to minimize the loss function $L(\Vec{t}, \Vec{\tau}) = \norm{\mathcal{U} -  \mathcal{V}(\Vec{t}, \Vec{\tau}) }^2$. Here $\norm{\cdot}$ denotes the Frobenius norm. Given access to the entries of a Haar random target unitary $\mathcal{U}$, we fix the number of parameters $2K$ and ask how many gradient descent steps $S$ are required to construct the sequence $\mathcal{V}(\Vec{t}, \Vec{\tau}) = e^{-i At_K} e^{-i B \tau _K}  \ldots e^{-i At_1} e^{-i B \tau _1} $ that can learn the target unitary $\mathcal{U}$ to a given accuracy or loss.
   
   We present numerical evidence that with at least $d^2$ parameters in the sequence $\mathcal{V}(\Vec{t}, \Vec{\tau})$, we can learn any selected Haar random unitary $\mathcal{U}$. Because of the highly non-convex nature of the loss landscape over the control parameters, we did not expect this result. The details of the numerical analysis are provided below. 
   
   We ran experiments for a Haar random target unitary $\mathcal{U}$ of dimension 32 while varying the $2K$ parameters in $\mathcal{V}(\Vec{t}, \Vec{\tau})$. At each step, we compute the gradient $\nabla_{\vec{t}, \vec{\tau}}{L}$ and perform gradient descent with fixed step size. 
   
   In Fig.(\ref{fig1}), we plot the loss $L(\vec{t}, \vec{\tau})$ as a function of the number of gradient descent steps $S$ for learning sequences $\mathcal{V}(\vec{t},\vec{\tau})$ of varying depth $K$. When the sequence $\mathcal{V}( \vec{t}, \vec{\tau})$ is  under-parameterized with $2K < d^2$ parameters, we find that the loss function $L(\vec{t}, \vec{\tau})$ initially decreases but then plateaus. Thus, in the under-parameterized loss landscape, we find that as expected, with high probability, the gradient descent algorithm reaches a sub-optimal value of the loss which cannot be decreased by further increasing the number of gradient descent steps. 
   
   When the number of parameters $2K$ in $\mathcal{V}( \vec{t}, \vec{\tau})$ is equal to $d^2$ or more, we find that gradient descent always converges to the target unitary -- there are apparently no sub-optimal local minima in the loss landscape. As noted above, this result was unexpected given the non-convex nature of the loss landscape. We also find that the rate of convergence grows with the degree of over-parameterization as shown in Fig.(\ref{fig1}). At the critical point where the number of parameters $2K=d^2$, we note the existence of a ``computational phase transition." At this critical point, the learning process converges to the desired target unitary, but the rate of convergence becomes very slow. For each parameter manifold of dimension $0.1 d^2 \leq 2 K \leq 2 d^2$, we performed ten experiments and each of the experiments has been plotted in Fig.(\ref{fig1}).
   
   
    \begin{figure}[t]
   \includegraphics[width=1\linewidth]{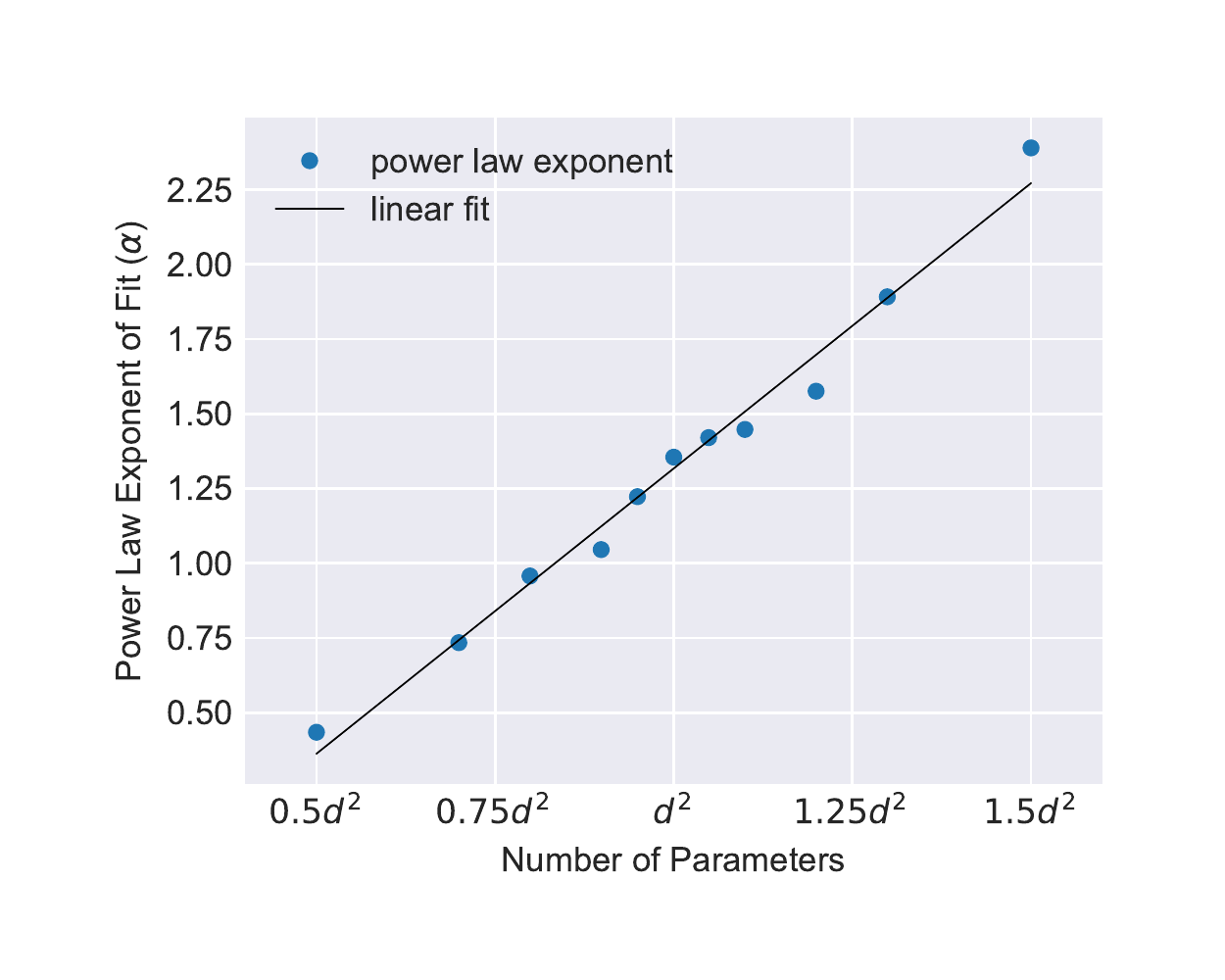}%
   \caption{The power law rate of gradient descent $\alpha$ for the initial 1000 gradient descent steps grows linearly with the number of parameters ($2K$). The first 50 steps have been excluded in the fit. The slope of the best fit line is $1.9$. The computational phase transition takes place at a value of $\alpha \approx 1.25$.}
   \label{power_law_linear}
   \end{figure}

   In Fig.(\ref{fig2}), we fit the loss $L(\vec{t},\vec{\tau})$ over the first 1000 gradient descent steps (the first 50 steps are excluded) to a power law 
   \begin{align}
   \label{eq:lossvsepochs}
   L = C_0 S^{-\alpha} + C_1,
   \end{align}
   where $C_0$ and $C_1$ are constants, $L = L(\vec{t}, \vec{\tau})$ and $S$ is the number of gradient descent steps.  As shown in Fig.(\ref{fig2}), the data for the initial 1000 gradient descent steps fits closely to such a power law. However, with the exception of the critical learning sequence with $2K = d^2$ parameters, the performance of gradient descent deviates from a power law fit at later steps. For the under-parameterized case, the gradient descent plateaus at a sub-optimal value of the loss. For the over-parameterized case, the power law transitions to an exponential as the gradient descent approaches the global minimum, which is consistent with the expected quadratic form of the loss function in the vicinity of the global minimum. Fig.(\ref{fig2}) shows the exponential fit for the later stages of gradient descent in the over-parameterized setting. The exponential fit takes the form
   
   \begin{align}
   \label{eq:lossvsepochs_expo}
   L = C_0 e^{-r(S-S_0)}+C_1,
   \end{align}
   where $C_0$, $C_1$, $r$, and $S_0$ are constants (optimized during the fit), $L = L(\vec{t}, \vec{\tau})$ and $S$ is the number of gradient descent steps.
   
   The critical case of the sequence $\mathcal{V}(\vec{t},\vec{\tau})$ with exactly $d^2$ parameters is consistent with a power law rate of convergence to the target unitary $\mathcal{U}$ during the entire gradient descent process.
   
   The initial power law form of the gradient descent is consistent with a loss landscape that obeys the relation $\Delta L / \Delta S \propto -S^{-(\alpha + 1)}$ and $\alpha \geq 0$. For example, the case $\alpha = 1$ corresponds to a power law of the form $\Delta L / \Delta S \propto -S^{-2}$. The final exponential form of convergence corresponds to the case $\alpha \to \infty$, and to a quadratic landscape where $\Delta L / \Delta S \propto -e^{-S} \propto -L$. The fitted value of $\alpha$ in the initial power law regime is plotted as a function of the number of parameters in Fig.(\ref{power_law_linear}). Here, we observe a linear relationship between the power law exponent $\alpha$ in Eq.~\eqref{eq:lossvsepochs} and the number of parameters $2K$ in $\mathcal{V}(\vec{t},\vec{\tau})$ -- \textit{i.e.}, the larger the degree of over-parameterization, the faster the rate of convergence, and the larger the exponent in the power law. 
   

   %
   \begin{figure}[t]
   \includegraphics[width=1\linewidth]{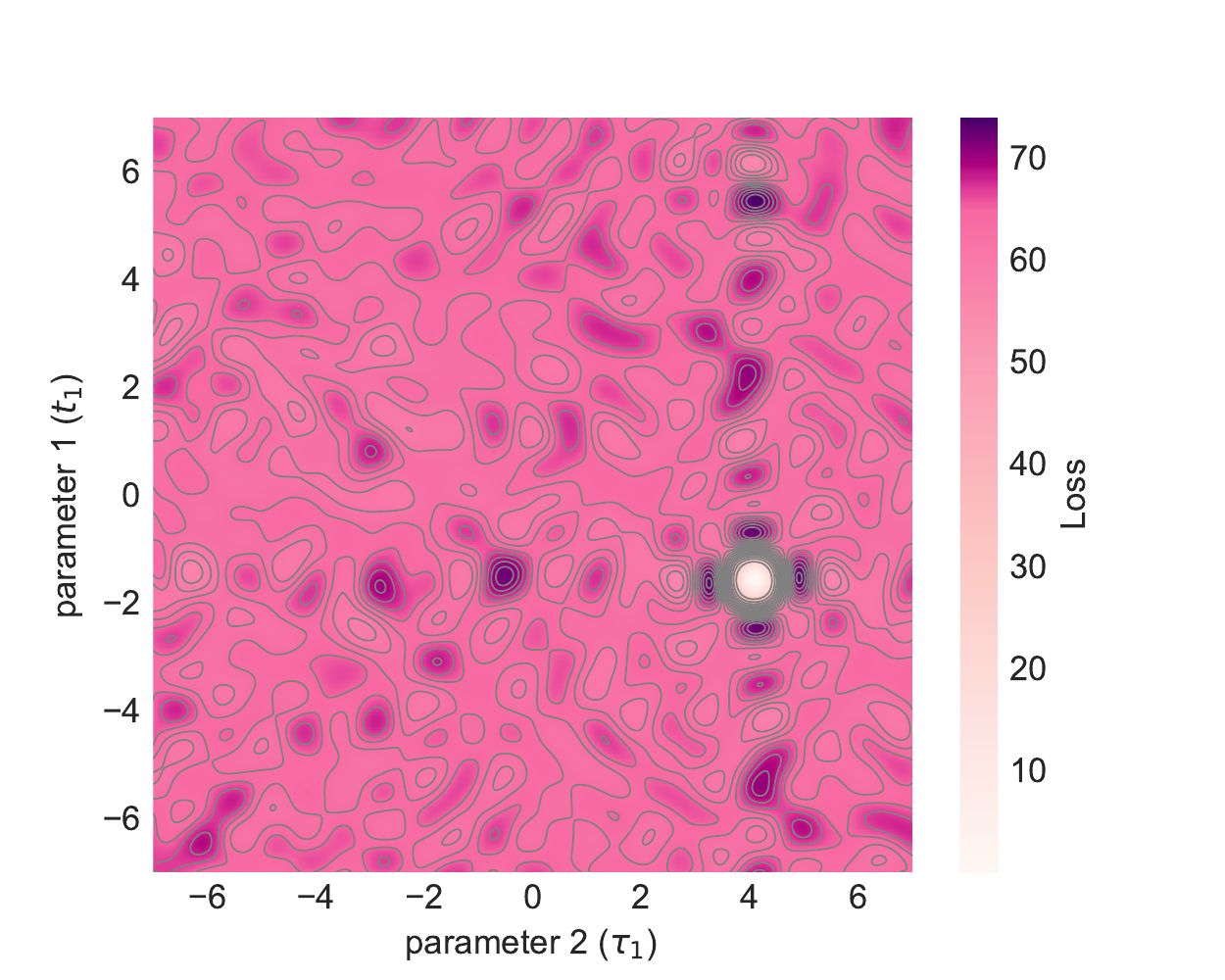}%
   \caption{Loss function landscape when the target unitary is $e^{-i A t^*}e^{-i B  \tau^*}$ where $t^* = -1.59 $ and $\tau^* = 4.08$. The landscape is highly non-convex with many local minima, indicating that it is difficult to learn the target unitary with first order optimization methods such as gradient descent unless the starting point of optimization lies in the neighbourhood of the global minimum.}
  \label{fig4}
   \end{figure}

   \section{Learning shallow-depth unitaries}
   
    In this section, we study the learnability of low-depth alternating operator unitaries $\mathcal{U}(\vec{t},\vec{\tau}) = e^{-i At_N} e^{-i B \tau _N}  \ldots e^{-i At_1} e^{-i B \tau _1}$ where $2N \ll d^2$. Such unitaries are the alternating operator analogue of shallow depth quantum circuits. As noted above, unitaries of this form are by definition, obtainable by a learning sequence $\mathcal{V}(\vec{t},\vec{\tau})$ with depth $K \geq N$. We wish to investigate for which values of $K$, it is possible to learn the target unitary $\mathcal{U}(\vec{t},\vec{\tau})$ of depth $N$. We could reasonably hope that such a shallow depth unitary could be learned by performing gradient descent over sequences $\mathcal{V}(\vec{t},\vec{\tau})$ of depth $K=N$. We find that this is not the case. Indeed, we find that even to learn a unitary $\mathcal{U}(\vec{t},\vec{\tau})$ of depth $N=1$, with high probability, we require a full depth learning sequence $\mathcal{V}(\vec{t},\vec{\tau})$ of depth $K \geq d^2/2$ or $2K \geq d^2$ parameters in $\mathcal{V}(\vec{t},\vec{\tau})$. 
    
     \begin{figure}[t]
      \includegraphics[width=1\linewidth]{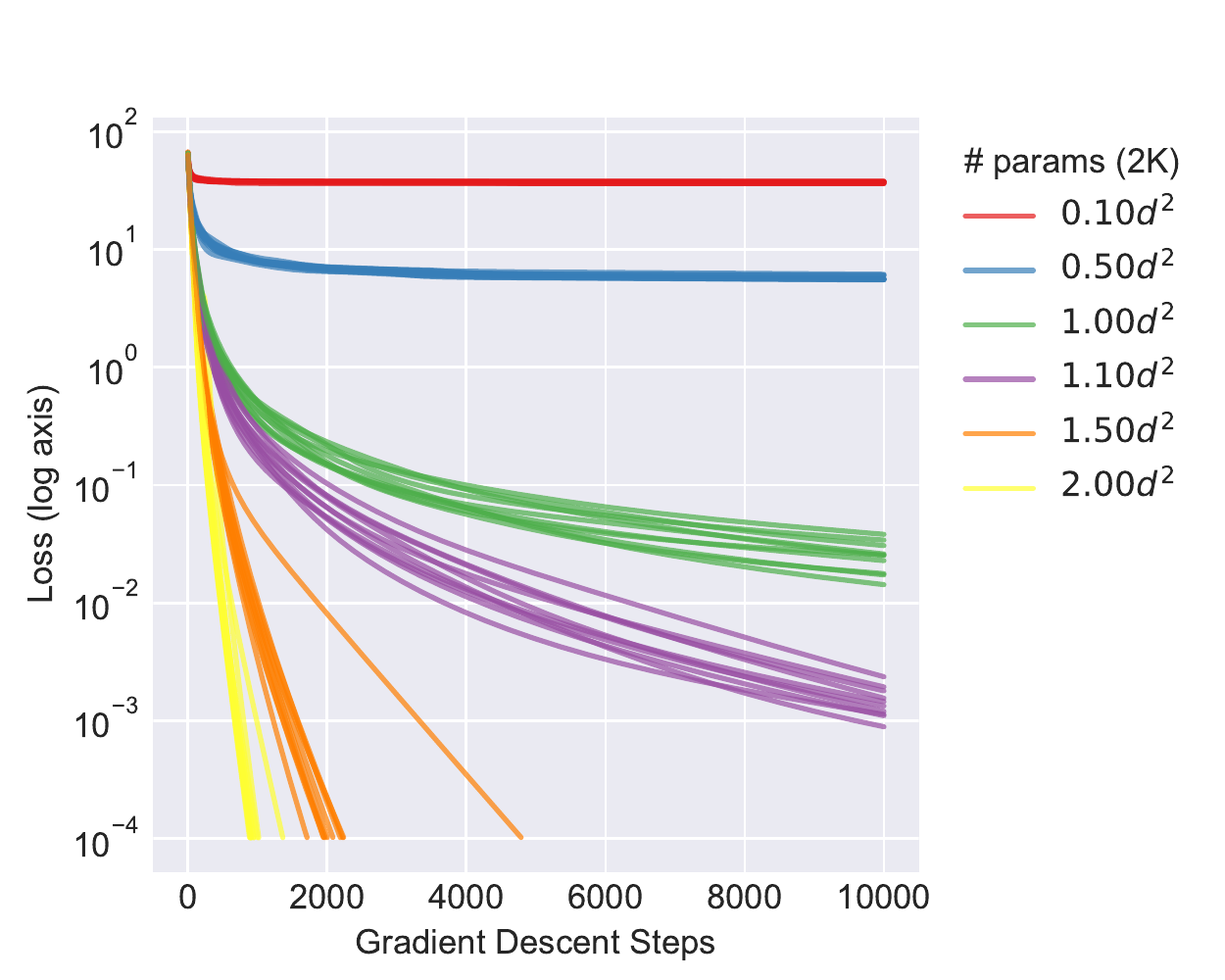}%
   \caption{Gradient descent experiments for a low-depth unitary $\mathcal{U}(t_1^*,\tau_1^*,t_2^*,\tau_2^*)$ of dimension 32 with 4 parameters ($N$=2) where $t_1^*,t_2^*,\tau_1^*,\tau_2^* \in [-2,2]$. }
   \label{fig5}
   \end{figure}
    
    Depth $N$=1 unitaries take the form $\mathcal{U}(t^*,\tau^*) = e^{-i A t^*}e^{-i B \tau^*}$. In Fig.(\ref{fig4}), we present the landscape of the loss function $L(t,\tau)= \norm{e^{-i A t^*}e^{-i B \tau^*} - e^{-i A t} e^{-i B \tau} }^2 $ which is a two dimensional parametric manifold. Here we attempt to learn the target unitary $\mathcal{U}(t^*,\tau^*)$ via a sequence $\mathcal{V}(t,\tau)$ also with two parameters. The loss function landscape is highly non-convex and contains many local sub-optimal traps. Learning the target unitary with much less than $d^2$ parameters using gradient descent is guaranteed only when the initial values of the parameters $t,\tau$ lie in the neighbourhood of the global minimum at $t^* = -1.59 $ and $\tau^* = 4.08$. In unbounded parametric manifolds, such an optimal initialization is generally hard to achieve.
    
    Next, we consider a target unitary $\mathcal{U}(\vec{t}^*,\vec{\tau}^*)$ with four parameters ($N=2$). In Fig.(\ref{fig5}), we find that when the sequence $\mathcal{V}(\vec{t},\vec{\tau})$ has $2 K < d^2$ parameters, the loss function plateaus with increasing gradient descent steps. This indicates that gradient descent halts at a local minimum of the loss function landscape. The rate of learning improves when $2K = d^2$ or $2K >  d^2$ as in the over-parameterized domain. In this setting, the loss function rapidly converges towards the global minimum of the landscape, and the rate of convergence to the target unitary $\mathcal{U}(\vec{t}^*,\vec{\tau}^*)$ is similar to the over-parameterized case shown in Fig.(\ref{fig2}). 
    
    Surjectivity in the map from control parameters to the tangent space of the unitary manifold has been shown to be a sufficient condition for constructing loss landscapes with no poor local minima in quantum control settings \cite{russell2016quantum}. This criteria implies that complete freedom of movement at any point in the unitary manifold is sufficient to guarantee convergence to a global minimum. The under-parameterized setting does not meet this criteria of surjectivity, since infinitesimal variations in the $2K<d^2$ parameters $\{ t_1, \tau_1, \dots, t_K, \tau_K \}$ are not sufficient to generate any local infinitesimal change in the unitary manifold of dimension $d^2$. When the number of control parameters is $d^2$ or greater, the map from controls to unitaries is locally surjective at almost all points of the control space, so that at almost all points, all directions in the space of unitaries can be obtained. Our numerical results suggest that when there are a sufficient number of control parameters to render the system controllable, the control map is locally surjective along the entire path of gradient descent all the way to the global optimum.

   \section{Conclusion}
   We have numerically analysed the hardness of obtaining the optimal control parameters in an alternating operator sequence for learning arbitrary unitaries using gradient descent optimization. For learning a Haar random target unitary in $d$ dimensions to a desired accuracy, we find that gradient descent requires at least $d^2$ parameters in an alternating operator sequence. When there are fewer than $d^2$ parameters in the sequence, gradient descent converges to an undesirable minimum of the loss function landscape which cannot be escaped with further gradient descent steps. This is true even for learning shallow-depth alternating operator target unitaries which are the alternating operator analogue of shallow depth quantum circuits. 
   
   Gradient descent methods generally guarantee convergence only in convex spaces. The loss function landscape for unitaries is highly non-convex, and when we began this investigation, we did not know whether gradient descent on $2K \geq d^2$ parameters in the landscape would succeed in the search for a global minimum. Indeed, we expected that gradient descent would not always converge. However, in contrast to our initial expectations, we find that when the number of parameters in the loss function landscape $2K \geq d^2$, gradient descent always converges to an optimal global minimum in the landscape. At the critical value of $2K = d^2$ parameters, we observe a ``computational phase transition" characterized by a power law convergence to the global optimum. 
   

   \section{Acknowledgements}
   We thank Milad Marvian, Giacomo de Palma, Zi-Wen Liu, Can Gokler, Dirk Englund, Herschel Rabitz and Yann LeCun for helpful discussions and suggestions. This work was  supported by DOE, IARPA, NSF, and ARO.
   
   \bibliography{QAOA.bib} 

\begin{thebibliography}{27}%
\makeatletter
\providecommand \@ifxundefined [1]{%
 \@ifx{#1\undefined}
}%
\providecommand \@ifnum [1]{%
 \ifnum #1\expandafter \@firstoftwo
 \else \expandafter \@secondoftwo
 \fi
}%
\providecommand \@ifx [1]{%
 \ifx #1\expandafter \@firstoftwo
 \else \expandafter \@secondoftwo
 \fi
}%
\providecommand \natexlab [1]{#1}%
\providecommand \enquote  [1]{``#1''}%
\providecommand \bibnamefont  [1]{#1}%
\providecommand \bibfnamefont [1]{#1}%
\providecommand \citenamefont [1]{#1}%
\providecommand \href@noop [0]{\@secondoftwo}%
\providecommand \href [0]{\begingroup \@sanitize@url \@href}%
\providecommand \@href[1]{\@@startlink{#1}\@@href}%
\providecommand \@@href[1]{\endgroup#1\@@endlink}%
\providecommand \@sanitize@url [0]{\catcode `\\12\catcode `\$12\catcode
  `\&12\catcode `\#12\catcode `\^12\catcode `\_12\catcode `\%12\relax}%
\providecommand \@@startlink[1]{}%
\providecommand \@@endlink[0]{}%
\providecommand \url  [0]{\begingroup\@sanitize@url \@url }%
\providecommand \@url [1]{\endgroup\@href {#1}{\urlprefix }}%
\providecommand \urlprefix  [0]{URL }%
\providecommand \Eprint [0]{\href }%
\providecommand \doibase [0]{http://dx.doi.org/}%
\providecommand \selectlanguage [0]{\@gobble}%
\providecommand \bibinfo  [0]{\@secondoftwo}%
\providecommand \bibfield  [0]{\@secondoftwo}%
\providecommand \translation [1]{[#1]}%
\providecommand \BibitemOpen [0]{}%
\providecommand \bibitemStop [0]{}%
\providecommand \bibitemNoStop [0]{.\EOS\space}%
\providecommand \EOS [0]{\spacefactor3000\relax}%
\providecommand \BibitemShut  [1]{\csname bibitem#1\endcsname}%
\let\auto@bib@innerbib\@empty
\bibitem [{\citenamefont {Farhi}\ \emph
  {et~al.}(2014{\natexlab{a}})\citenamefont {Farhi}, \citenamefont
  {Goldstone},\ and\ \citenamefont {Gutmann}}]{farhi2014quantum1}%
  \BibitemOpen
  \bibfield  {author} {\bibinfo {author} {\bibfnamefont {E.}~\bibnamefont
  {Farhi}}, \bibinfo {author} {\bibfnamefont {J.}~\bibnamefont {Goldstone}}, \
  and\ \bibinfo {author} {\bibfnamefont {S.}~\bibnamefont {Gutmann}},\
  }\bibfield  {title} {\enquote {\bibinfo {title} {A quantum approximate
  optimization algorithm},}\ }\href@noop {} {\bibfield  {journal} {\bibinfo
  {journal} {arXiv:1411.4028}\ } (\bibinfo {year}
  {2014}{\natexlab{a}})}\BibitemShut {NoStop}%
\bibitem [{\citenamefont {Farhi}\ \emph
  {et~al.}(2014{\natexlab{b}})\citenamefont {Farhi}, \citenamefont
  {Goldstone},\ and\ \citenamefont {Gutmann}}]{farhi2014quantum2}%
  \BibitemOpen
  \bibfield  {author} {\bibinfo {author} {\bibfnamefont {E.}~\bibnamefont
  {Farhi}}, \bibinfo {author} {\bibfnamefont {J.}~\bibnamefont {Goldstone}}, \
  and\ \bibinfo {author} {\bibfnamefont {S.}~\bibnamefont {Gutmann}},\
  }\bibfield  {title} {\enquote {\bibinfo {title} {A quantum approximate
  optimization algorithm applied to a bounded occurrence constraint problem},}\
  }\href@noop {} {\bibfield  {journal} {\bibinfo  {journal} {arXiv:1412.6062}\
  } (\bibinfo {year} {2014}{\natexlab{b}})}\BibitemShut {NoStop}%
\bibitem [{\citenamefont {Lloyd}(2018)}]{lloyd2018quantum}%
  \BibitemOpen
  \bibfield  {author} {\bibinfo {author} {\bibfnamefont {S.}~\bibnamefont
  {Lloyd}},\ }\bibfield  {title} {\enquote {\bibinfo {title} {Quantum
  approximate optimization is computationally universal},}\ }\href@noop {}
  {\bibfield  {journal} {\bibinfo  {journal} {arXiv:1812.11075}\ } (\bibinfo
  {year} {2018})}\BibitemShut {NoStop}%
\bibitem [{\citenamefont {Rabitz}\ \emph {et~al.}(2004)\citenamefont {Rabitz},
  \citenamefont {Hsieh},\ and\ \citenamefont {Rosenthal}}]{rabitz2004quantum}%
  \BibitemOpen
  \bibfield  {author} {\bibinfo {author} {\bibfnamefont {H.}~\bibnamefont
  {Rabitz}}, \bibinfo {author} {\bibfnamefont {M.}~\bibnamefont {Hsieh}}, \
  and\ \bibinfo {author} {\bibfnamefont {C.}~\bibnamefont {Rosenthal}},\
  }\bibfield  {title} {\enquote {\bibinfo {title} {Quantum optimally controlled
  transition landscapes},}\ }\href@noop {} {\bibfield  {journal} {\bibinfo
  {journal} {Science}\ }\textbf {\bibinfo {volume} {303}},\ \bibinfo {pages}
  {1998--2001} (\bibinfo {year} {2004})}\BibitemShut {NoStop}%
\bibitem [{\citenamefont {Khaneja}\ \emph {et~al.}(2005)\citenamefont
  {Khaneja}, \citenamefont {Reiss}, \citenamefont {Kehlet}, \citenamefont
  {Schulte-Herbr{\"u}ggen},\ and\ \citenamefont {Glaser}}]{khaneja2005optimal}%
  \BibitemOpen
  \bibfield  {author} {\bibinfo {author} {\bibfnamefont {N.}~\bibnamefont
  {Khaneja}}, \bibinfo {author} {\bibfnamefont {T.}~\bibnamefont {Reiss}},
  \bibinfo {author} {\bibfnamefont {C.}~\bibnamefont {Kehlet}}, \bibinfo
  {author} {\bibfnamefont {T.}~\bibnamefont {Schulte-Herbr{\"u}ggen}}, \ and\
  \bibinfo {author} {\bibfnamefont {S.~J.}\ \bibnamefont {Glaser}},\ }\bibfield
   {title} {\enquote {\bibinfo {title} {Optimal control of coupled spin
  dynamics: design of \uppercase{N}\uppercase{M}\uppercase{R} pulse sequences
  by gradient ascent algorithms},}\ }\href@noop {} {\bibfield  {journal}
  {\bibinfo  {journal} {Journal of Magnetic Resonance}\ }\textbf {\bibinfo
  {volume} {172}},\ \bibinfo {pages} {296--305} (\bibinfo {year}
  {2005})}\BibitemShut {NoStop}%
\bibitem [{\citenamefont {Rabitz}\ \emph {et~al.}(2005)\citenamefont {Rabitz},
  \citenamefont {Hsieh},\ and\ \citenamefont
  {Rosenthal}}]{rabitz2005landscape}%
  \BibitemOpen
  \bibfield  {author} {\bibinfo {author} {\bibfnamefont {H.}~\bibnamefont
  {Rabitz}}, \bibinfo {author} {\bibfnamefont {M.}~\bibnamefont {Hsieh}}, \
  and\ \bibinfo {author} {\bibfnamefont {C.}~\bibnamefont {Rosenthal}},\
  }\bibfield  {title} {\enquote {\bibinfo {title} {Landscape for optimal
  control of quantum-mechanical unitary transformations},}\ }\href@noop {}
  {\bibfield  {journal} {\bibinfo  {journal} {Physical Review A}\ }\textbf
  {\bibinfo {volume} {72}},\ \bibinfo {pages} {052337} (\bibinfo {year}
  {2005})}\BibitemShut {NoStop}%
\bibitem [{\citenamefont {Rabitz}\ \emph {et~al.}(2006)\citenamefont {Rabitz},
  \citenamefont {Ho}, \citenamefont {Hsieh}, \citenamefont {Kosut},\ and\
  \citenamefont {Demiralp}}]{rabitz2006topology}%
  \BibitemOpen
  \bibfield  {author} {\bibinfo {author} {\bibfnamefont {H.}~\bibnamefont
  {Rabitz}}, \bibinfo {author} {\bibfnamefont {T.-S.}\ \bibnamefont {Ho}},
  \bibinfo {author} {\bibfnamefont {M.}~\bibnamefont {Hsieh}}, \bibinfo
  {author} {\bibfnamefont {R.}~\bibnamefont {Kosut}}, \ and\ \bibinfo {author}
  {\bibfnamefont {M.}~\bibnamefont {Demiralp}},\ }\bibfield  {title} {\enquote
  {\bibinfo {title} {Topology of optimally controlled quantum mechanical
  transition probability landscapes},}\ }\href@noop {} {\bibfield  {journal}
  {\bibinfo  {journal} {Physical Review A}\ }\textbf {\bibinfo {volume} {74}},\
  \bibinfo {pages} {012721} (\bibinfo {year} {2006})}\BibitemShut {NoStop}%
\bibitem [{\citenamefont {Chakrabarti}\ and\ \citenamefont
  {Rabitz}(2007)}]{chakrabarti2007quantum}%
  \BibitemOpen
  \bibfield  {author} {\bibinfo {author} {\bibfnamefont {R.}~\bibnamefont
  {Chakrabarti}}\ and\ \bibinfo {author} {\bibfnamefont {H.}~\bibnamefont
  {Rabitz}},\ }\bibfield  {title} {\enquote {\bibinfo {title} {Quantum control
  landscapes},}\ }\href@noop {} {\bibfield  {journal} {\bibinfo  {journal}
  {International Reviews in Physical Chemistry}\ }\textbf {\bibinfo {volume}
  {26}},\ \bibinfo {pages} {671--735} (\bibinfo {year} {2007})}\BibitemShut
  {NoStop}%
\bibitem [{\citenamefont {Moore}\ \emph {et~al.}(2008)\citenamefont {Moore},
  \citenamefont {Hsieh},\ and\ \citenamefont {Rabitz}}]{moore2008relationship}%
  \BibitemOpen
  \bibfield  {author} {\bibinfo {author} {\bibfnamefont {K.}~\bibnamefont
  {Moore}}, \bibinfo {author} {\bibfnamefont {M.}~\bibnamefont {Hsieh}}, \ and\
  \bibinfo {author} {\bibfnamefont {H.}~\bibnamefont {Rabitz}},\ }\bibfield
  {title} {\enquote {\bibinfo {title} {On the relationship between quantum
  control landscape structure and optimization complexity},}\ }\href@noop {}
  {\bibfield  {journal} {\bibinfo  {journal} {The Journal of chemical physics}\
  }\textbf {\bibinfo {volume} {128}},\ \bibinfo {pages} {154117} (\bibinfo
  {year} {2008})}\BibitemShut {NoStop}%
\bibitem [{\citenamefont {Ho}\ \emph {et~al.}(2009)\citenamefont {Ho},
  \citenamefont {Dominy},\ and\ \citenamefont {Rabitz}}]{rabitz2009landscape}%
  \BibitemOpen
  \bibfield  {author} {\bibinfo {author} {\bibfnamefont {T.-S.}\ \bibnamefont
  {Ho}}, \bibinfo {author} {\bibfnamefont {J.}~\bibnamefont {Dominy}}, \ and\
  \bibinfo {author} {\bibfnamefont {H.}~\bibnamefont {Rabitz}},\ }\bibfield
  {title} {\enquote {\bibinfo {title} {Landscape of unitary transformations in
  controlled quantum dynamics},}\ }\href@noop {} {\bibfield  {journal}
  {\bibinfo  {journal} {Physical Review A}\ }\textbf {\bibinfo {volume} {79}},\
  \bibinfo {pages} {013422} (\bibinfo {year} {2009})}\BibitemShut {NoStop}%
\bibitem [{\citenamefont {Brif}\ \emph {et~al.}(2010)\citenamefont {Brif},
  \citenamefont {Chakrabarti},\ and\ \citenamefont {Rabitz}}]{brif2010control}%
  \BibitemOpen
  \bibfield  {author} {\bibinfo {author} {\bibfnamefont {C.}~\bibnamefont
  {Brif}}, \bibinfo {author} {\bibfnamefont {R.}~\bibnamefont {Chakrabarti}}, \
  and\ \bibinfo {author} {\bibfnamefont {H.}~\bibnamefont {Rabitz}},\
  }\bibfield  {title} {\enquote {\bibinfo {title} {Control of quantum
  phenomena: past, present and future},}\ }\href@noop {} {\bibfield  {journal}
  {\bibinfo  {journal} {New Journal of Physics}\ }\textbf {\bibinfo {volume}
  {12}},\ \bibinfo {pages} {075008} (\bibinfo {year} {2010})}\BibitemShut
  {NoStop}%
\bibitem [{\citenamefont {Riviello}\ \emph {et~al.}(2015)\citenamefont
  {Riviello}, \citenamefont {Tibbetts}, \citenamefont {Brif}, \citenamefont
  {Long}, \citenamefont {Wu}, \citenamefont {Ho},\ and\ \citenamefont
  {Rabitz}}]{riviello2015searching}%
  \BibitemOpen
  \bibfield  {author} {\bibinfo {author} {\bibfnamefont {G.}~\bibnamefont
  {Riviello}}, \bibinfo {author} {\bibfnamefont {K.~M.}\ \bibnamefont
  {Tibbetts}}, \bibinfo {author} {\bibfnamefont {C.}~\bibnamefont {Brif}},
  \bibinfo {author} {\bibfnamefont {R.}~\bibnamefont {Long}}, \bibinfo {author}
  {\bibfnamefont {R.-B.}\ \bibnamefont {Wu}}, \bibinfo {author} {\bibfnamefont
  {T.-S.}\ \bibnamefont {Ho}}, \ and\ \bibinfo {author} {\bibfnamefont
  {H.}~\bibnamefont {Rabitz}},\ }\bibfield  {title} {\enquote {\bibinfo {title}
  {Searching for quantum optimal controls under severe constraints},}\
  }\href@noop {} {\bibfield  {journal} {\bibinfo  {journal} {Physical Review
  A}\ }\textbf {\bibinfo {volume} {91}},\ \bibinfo {pages} {043401} (\bibinfo
  {year} {2015})}\BibitemShut {NoStop}%
\bibitem [{\citenamefont {Riviello}\ \emph {et~al.}(2017)\citenamefont
  {Riviello}, \citenamefont {Wu}, \citenamefont {Sun},\ and\ \citenamefont
  {Rabitz}}]{riviello2017searching}%
  \BibitemOpen
  \bibfield  {author} {\bibinfo {author} {\bibfnamefont {G.}~\bibnamefont
  {Riviello}}, \bibinfo {author} {\bibfnamefont {R.-B.}\ \bibnamefont {Wu}},
  \bibinfo {author} {\bibfnamefont {Q.}~\bibnamefont {Sun}}, \ and\ \bibinfo
  {author} {\bibfnamefont {H.}~\bibnamefont {Rabitz}},\ }\bibfield  {title}
  {\enquote {\bibinfo {title} {Searching for an optimal control in the presence
  of saddles on the quantum-mechanical observable landscape},}\ }\href@noop {}
  {\bibfield  {journal} {\bibinfo  {journal} {Physical Review A}\ }\textbf
  {\bibinfo {volume} {95}},\ \bibinfo {pages} {063418} (\bibinfo {year}
  {2017})}\BibitemShut {NoStop}%
\bibitem [{\citenamefont {Russell}\ \emph {et~al.}(2017)\citenamefont
  {Russell}, \citenamefont {Rabitz},\ and\ \citenamefont
  {Wu}}]{russell2016quantum}%
  \BibitemOpen
  \bibfield  {author} {\bibinfo {author} {\bibfnamefont {B.}~\bibnamefont
  {Russell}}, \bibinfo {author} {\bibfnamefont {H.}~\bibnamefont {Rabitz}}, \
  and\ \bibinfo {author} {\bibfnamefont {R.-B.}\ \bibnamefont {Wu}},\
  }\bibfield  {title} {\enquote {\bibinfo {title} {Quantum control landscapes
  are almost always trap free},}\ }\href@noop {} {\bibfield  {journal}
  {\bibinfo  {journal} {Journal of Physics A: Mathematical and Theoretical}\
  }\textbf {\bibinfo {volume} {50}},\ \bibinfo {pages} {205302} (\bibinfo
  {year} {2017})}\BibitemShut {NoStop}%
\bibitem [{\citenamefont {Lloyd}\ and\ \citenamefont
  {Maity}(2019)}]{lloyd2019efficient}%
  \BibitemOpen
  \bibfield  {author} {\bibinfo {author} {\bibfnamefont {S.}~\bibnamefont
  {Lloyd}}\ and\ \bibinfo {author} {\bibfnamefont {R.}~\bibnamefont {Maity}},\
  }\bibfield  {title} {\enquote {\bibinfo {title} {Efficient implementation of
  unitary transformations},}\ }\href@noop {} {\bibfield  {journal} {\bibinfo
  {journal} {arXiv:1901.03431}\ } (\bibinfo {year} {2019})}\BibitemShut
  {NoStop}%
\bibitem [{\citenamefont {Farhi}\ and\ \citenamefont
  {Harrow}(2016)}]{farhi2016quantum}%
  \BibitemOpen
  \bibfield  {author} {\bibinfo {author} {\bibfnamefont {E.}~\bibnamefont
  {Farhi}}\ and\ \bibinfo {author} {\bibfnamefont {A.~W.}\ \bibnamefont
  {Harrow}},\ }\bibfield  {title} {\enquote {\bibinfo {title} {Quantum
  supremacy through the quantum approximate optimization algorithm},}\
  }\href@noop {} {\bibfield  {journal} {\bibinfo  {journal} {arXiv:1602.07674}\
  } (\bibinfo {year} {2016})}\BibitemShut {NoStop}%
\bibitem [{\citenamefont {Jiang}\ \emph {et~al.}(2017)\citenamefont {Jiang},
  \citenamefont {Rieffel},\ and\ \citenamefont {Wang}}]{jiang2017qaoa}%
  \BibitemOpen
  \bibfield  {author} {\bibinfo {author} {\bibfnamefont {Z.}~\bibnamefont
  {Jiang}}, \bibinfo {author} {\bibfnamefont {E.}~\bibnamefont {Rieffel}}, \
  and\ \bibinfo {author} {\bibfnamefont {Z.}~\bibnamefont {Wang}},\ }\bibfield
  {title} {\enquote {\bibinfo {title} {A
  \uppercase{Q}\uppercase{A}\uppercase{O}\uppercase{A}-inspired circuit for
  \uppercase{G}rover’s unstructured search using a transverse field},}\
  }\href@noop {} {\bibfield  {journal} {\bibinfo  {journal} {arXiv:1702.0257}\
  } (\bibinfo {year} {2017})}\BibitemShut {NoStop}%
\bibitem [{\citenamefont {Zhou}\ \emph {et~al.}(2018)\citenamefont {Zhou},
  \citenamefont {Wang}, \citenamefont {Choi}, \citenamefont {Pichler},\ and\
  \citenamefont {Lukin}}]{zhou2018quantum}%
  \BibitemOpen
  \bibfield  {author} {\bibinfo {author} {\bibfnamefont {L.}~\bibnamefont
  {Zhou}}, \bibinfo {author} {\bibfnamefont {S.-T.}\ \bibnamefont {Wang}},
  \bibinfo {author} {\bibfnamefont {S.}~\bibnamefont {Choi}}, \bibinfo {author}
  {\bibfnamefont {H.}~\bibnamefont {Pichler}}, \ and\ \bibinfo {author}
  {\bibfnamefont {M.~D.}\ \bibnamefont {Lukin}},\ }\bibfield  {title} {\enquote
  {\bibinfo {title} {Quantum approximate optimization algorithm: performance,
  mechanism, and implementation on near-term devices},}\ }\href@noop {}
  {\bibfield  {journal} {\bibinfo  {journal} {arXiv:1812.01041}\ } (\bibinfo
  {year} {2018})}\BibitemShut {NoStop}%
\bibitem [{\citenamefont {Gily{\'e}n}\ \emph {et~al.}(2019)\citenamefont
  {Gily{\'e}n}, \citenamefont {Arunachalam},\ and\ \citenamefont
  {Wiebe}}]{gilyen2019optimizing}%
  \BibitemOpen
  \bibfield  {author} {\bibinfo {author} {\bibfnamefont {A.}~\bibnamefont
  {Gily{\'e}n}}, \bibinfo {author} {\bibfnamefont {S.}~\bibnamefont
  {Arunachalam}}, \ and\ \bibinfo {author} {\bibfnamefont {N.}~\bibnamefont
  {Wiebe}},\ }\bibfield  {title} {\enquote {\bibinfo {title} {Optimizing
  quantum optimization algorithms via faster quantum gradient computation},}\
  }in\ \href@noop {} {\emph {\bibinfo {booktitle} {Proceedings of the Thirtieth
  Annual ACM-SIAM Symposium on Discrete Algorithms}}}\ (\bibinfo {organization}
  {SIAM},\ \bibinfo {year} {2019})\ pp.\ \bibinfo {pages}
  {1425--1444}\BibitemShut {NoStop}%
\bibitem [{\citenamefont {McClean}\ \emph {et~al.}(2016)\citenamefont
  {McClean}, \citenamefont {Romero}, \citenamefont {Babbush},\ and\
  \citenamefont {Aspuru-Guzik}}]{mcclean2016theory}%
  \BibitemOpen
  \bibfield  {author} {\bibinfo {author} {\bibfnamefont {J.~R.}\ \bibnamefont
  {McClean}}, \bibinfo {author} {\bibfnamefont {J.}~\bibnamefont {Romero}},
  \bibinfo {author} {\bibfnamefont {R.}~\bibnamefont {Babbush}}, \ and\
  \bibinfo {author} {\bibfnamefont {A.}~\bibnamefont {Aspuru-Guzik}},\
  }\bibfield  {title} {\enquote {\bibinfo {title} {The theory of variational
  hybrid quantum-classical algorithms},}\ }\href@noop {} {\bibfield  {journal}
  {\bibinfo  {journal} {New Journal of Physics}\ }\textbf {\bibinfo {volume}
  {18}},\ \bibinfo {pages} {023023} (\bibinfo {year} {2016})}\BibitemShut
  {NoStop}%
\bibitem [{\citenamefont {Peruzzo}\ \emph {et~al.}(2014)\citenamefont
  {Peruzzo}, \citenamefont {McClean}, \citenamefont {Shadbolt}, \citenamefont
  {Yung}, \citenamefont {Zhou}, \citenamefont {Love}, \citenamefont
  {Aspuru-Guzik},\ and\ \citenamefont {O’brien}}]{peruzzo2014variational}%
  \BibitemOpen
  \bibfield  {author} {\bibinfo {author} {\bibfnamefont {A.}~\bibnamefont
  {Peruzzo}}, \bibinfo {author} {\bibfnamefont {J.}~\bibnamefont {McClean}},
  \bibinfo {author} {\bibfnamefont {P.}~\bibnamefont {Shadbolt}}, \bibinfo
  {author} {\bibfnamefont {M.-H.}\ \bibnamefont {Yung}}, \bibinfo {author}
  {\bibfnamefont {X.-Q.}\ \bibnamefont {Zhou}}, \bibinfo {author}
  {\bibfnamefont {P.~J.}\ \bibnamefont {Love}}, \bibinfo {author}
  {\bibfnamefont {A.}~\bibnamefont {Aspuru-Guzik}}, \ and\ \bibinfo {author}
  {\bibfnamefont {J.}~\bibnamefont {O’brien}},\ }\bibfield  {title} {\enquote
  {\bibinfo {title} {A variational eigenvalue solver on a photonic quantum
  processor},}\ }\href@noop {} {\bibfield  {journal} {\bibinfo  {journal}
  {Nature communications}\ }\textbf {\bibinfo {volume} {5}},\ \bibinfo {pages}
  {4213} (\bibinfo {year} {2014})}\BibitemShut {NoStop}%
\bibitem [{\citenamefont {Khatri}\ \emph {et~al.}(2019)\citenamefont {Khatri},
  \citenamefont {LaRose}, \citenamefont {Poremba}, \citenamefont {Cincio},
  \citenamefont {Sornborger},\ and\ \citenamefont
  {Coles}}]{khatri2019quantumassisted}%
  \BibitemOpen
  \bibfield  {author} {\bibinfo {author} {\bibfnamefont {S.}~\bibnamefont
  {Khatri}}, \bibinfo {author} {\bibfnamefont {R.}~\bibnamefont {LaRose}},
  \bibinfo {author} {\bibfnamefont {A.}~\bibnamefont {Poremba}}, \bibinfo
  {author} {\bibfnamefont {L.}~\bibnamefont {Cincio}}, \bibinfo {author}
  {\bibfnamefont {A.~T.}\ \bibnamefont {Sornborger}}, \ and\ \bibinfo {author}
  {\bibfnamefont {P.~J.}\ \bibnamefont {Coles}},\ }\bibfield  {title} {\enquote
  {\bibinfo {title} {Quantum-assisted quantum compiling},}\ }\href@noop {}
  {\bibfield  {journal} {\bibinfo  {journal} {{Quantum}}\ }\textbf {\bibinfo
  {volume} {3}},\ \bibinfo {pages} {140} (\bibinfo {year} {2019})}\BibitemShut
  {NoStop}%
\bibitem [{\citenamefont {Sharma}\ \emph {et~al.}()\citenamefont {Sharma},
  \citenamefont {Khatri}, \citenamefont {Cerezo},\ and\ \citenamefont
  {Coles}}]{sharma2019noise}%
  \BibitemOpen
  \bibfield  {author} {\bibinfo {author} {\bibfnamefont {K.}~\bibnamefont
  {Sharma}}, \bibinfo {author} {\bibfnamefont {S.}~\bibnamefont {Khatri}},
  \bibinfo {author} {\bibfnamefont {M.}~\bibnamefont {Cerezo}}, \ and\ \bibinfo
  {author} {\bibfnamefont {P.~J.}\ \bibnamefont {Coles}},\ }\bibfield  {title}
  {\enquote {\bibinfo {title} {Noise resilience of variational quantum
  compiling},}\ }\href@noop {} {\bibinfo  {journal} {arXiv:1908.04416
  [quant-ph]}\ }\BibitemShut {NoStop}%
\bibitem [{\citenamefont {Carolan}\ \emph {et~al.}(2020)\citenamefont
  {Carolan}, \citenamefont {Mohseni}, \citenamefont {Olson}, \citenamefont
  {Prabhu}, \citenamefont {Chen}, \citenamefont {Bunandar}, \citenamefont
  {Niu}, \citenamefont {Harris}, \citenamefont {Wong}, \citenamefont {Hochberg}
  \emph {et~al.}}]{carolan2020variational}%
  \BibitemOpen
\bibfield  {journal} {  }\bibfield  {author} {\bibinfo {author} {\bibfnamefont
  {J.}~\bibnamefont {Carolan}}, \bibinfo {author} {\bibfnamefont
  {M.}~\bibnamefont {Mohseni}}, \bibinfo {author} {\bibfnamefont {J.~P.}\
  \bibnamefont {Olson}}, \bibinfo {author} {\bibfnamefont {M.}~\bibnamefont
  {Prabhu}}, \bibinfo {author} {\bibfnamefont {C.}~\bibnamefont {Chen}},
  \bibinfo {author} {\bibfnamefont {D.}~\bibnamefont {Bunandar}}, \bibinfo
  {author} {\bibfnamefont {M.~Y.}\ \bibnamefont {Niu}}, \bibinfo {author}
  {\bibfnamefont {N.~C.}\ \bibnamefont {Harris}}, \bibinfo {author}
  {\bibfnamefont {F.}~\bibnamefont {Wong}}, \bibinfo {author} {\bibfnamefont
  {M.}~\bibnamefont {Hochberg}},  \emph {et~al.},\ }\bibfield  {title}
  {\enquote {\bibinfo {title} {Variational quantum unsampling on a quantum
  photonic processor},}\ }\href@noop {} {\bibfield  {journal} {\bibinfo
  {journal} {Nature Physics}\ ,\ \bibinfo {pages} {1--6}} (\bibinfo {year}
  {2020})}\BibitemShut {NoStop}%
\bibitem [{\citenamefont {Zaheer}\ \emph {et~al.}(2018)\citenamefont {Zaheer},
  \citenamefont {Reddi}, \citenamefont {Sachan}, \citenamefont {Kale},\ and\
  \citenamefont {Kumar}}]{zaheer2018adaptive}%
  \BibitemOpen
  \bibfield  {author} {\bibinfo {author} {\bibfnamefont {M.}~\bibnamefont
  {Zaheer}}, \bibinfo {author} {\bibfnamefont {S.}~\bibnamefont {Reddi}},
  \bibinfo {author} {\bibfnamefont {D.}~\bibnamefont {Sachan}}, \bibinfo
  {author} {\bibfnamefont {S.}~\bibnamefont {Kale}}, \ and\ \bibinfo {author}
  {\bibfnamefont {S.}~\bibnamefont {Kumar}},\ }\bibfield  {title} {\enquote
  {\bibinfo {title} {Adaptive methods for nonconvex optimization},}\ }in\
  \href@noop {} {\emph {\bibinfo {booktitle} {Advances in Neural Information
  Processing Systems}}}\ (\bibinfo {year} {2018})\ pp.\ \bibinfo {pages}
  {9793--9803}\BibitemShut {NoStop}%
\bibitem [{\citenamefont {Kingma}\ and\ \citenamefont
  {Ba}(2014)}]{kingma2014adam}%
  \BibitemOpen
  \bibfield  {author} {\bibinfo {author} {\bibfnamefont {D.~P.}\ \bibnamefont
  {Kingma}}\ and\ \bibinfo {author} {\bibfnamefont {J.}~\bibnamefont {Ba}},\
  }\bibfield  {title} {\enquote {\bibinfo {title} {Adam: {A} method for
  stochastic optimization},}\ }\bibfield  {booktitle} {\emph {\bibinfo
  {booktitle} {3rd International Conference on Learning Representations, {ICLR}
  2015}},\ }\href@noop {} {\bibfield  {journal} {\bibinfo  {journal}
  {arXiv:1412.6980}\ } (\bibinfo {year} {2014})}\BibitemShut {NoStop}%
\bibitem [{\citenamefont {Paszke}\ \emph {et~al.}(2019)\citenamefont {Paszke},
  \citenamefont {Gross}, \citenamefont {Massa}, \citenamefont {Lerer},
  \citenamefont {Bradbury}, \citenamefont {Chanan}, \citenamefont {Killeen},
  \citenamefont {Lin}, \citenamefont {Gimelshein}, \citenamefont {Antiga} \emph
  {et~al.}}]{NEURIPS2019_9015}%
  \BibitemOpen
  \bibfield  {author} {\bibinfo {author} {\bibfnamefont {A.}~\bibnamefont
  {Paszke}}, \bibinfo {author} {\bibfnamefont {S.}~\bibnamefont {Gross}},
  \bibinfo {author} {\bibfnamefont {F.}~\bibnamefont {Massa}}, \bibinfo
  {author} {\bibfnamefont {A.}~\bibnamefont {Lerer}}, \bibinfo {author}
  {\bibfnamefont {J.}~\bibnamefont {Bradbury}}, \bibinfo {author}
  {\bibfnamefont {G.}~\bibnamefont {Chanan}}, \bibinfo {author} {\bibfnamefont
  {T.}~\bibnamefont {Killeen}}, \bibinfo {author} {\bibfnamefont
  {Z.}~\bibnamefont {Lin}}, \bibinfo {author} {\bibfnamefont {N.}~\bibnamefont
  {Gimelshein}}, \bibinfo {author} {\bibfnamefont {L.}~\bibnamefont {Antiga}},
  \emph {et~al.},\ }\bibfield  {title} {\enquote {\bibinfo {title} {Pytorch: An
  imperative style, high-performance deep learning library},}\ }in\ \href@noop
  {} {\emph {\bibinfo {booktitle} {Advances in Neural Information Processing
  Systems 32}}}\ (\bibinfo {year} {2019})\ pp.\ \bibinfo {pages}
  {8024--8035}\BibitemShut {NoStop}%
\end{thebibliography}%


\pagebreak
\widetext
\newpage

\begin{center}
\textbf{\large Supplemental Materials}
\end{center}
\setcounter{equation}{0}
\setcounter{figure}{0}
\setcounter{table}{0}
\setcounter{page}{1}
\makeatletter
\renewcommand{\theequation}{S\arabic{equation}}
\renewcommand{\thefigure}{S\arabic{figure}}
\renewcommand{\bibnumfmt}[1]{[S#1]}
\renewcommand{\citenumfont}[1]{S#1}

\section{Experiments using Adam optimizer}

\begin{figure}[h]
\subfloat[]{%
  \includegraphics[width=0.5\linewidth]{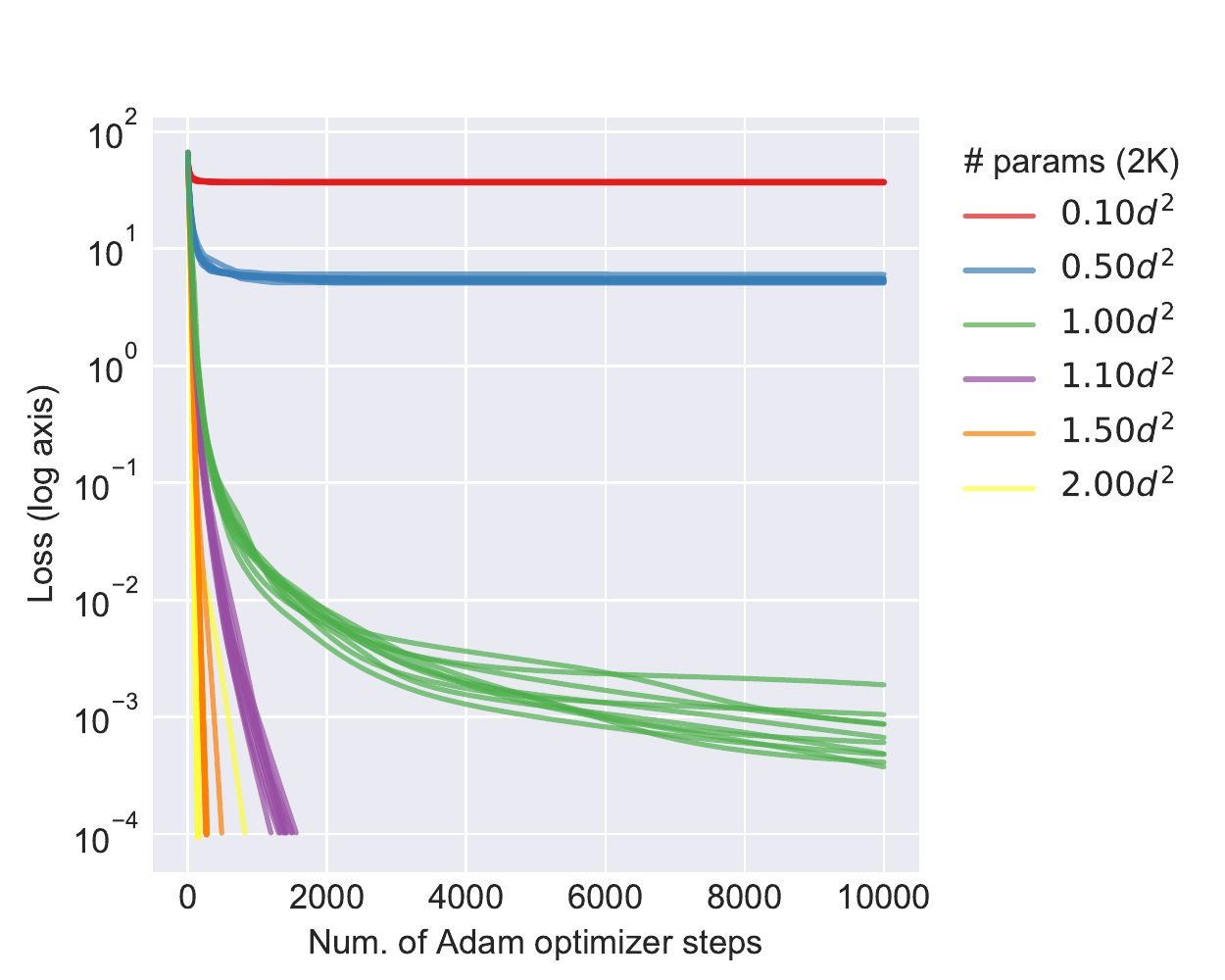}%
}\hfill
\subfloat[]{%
  \includegraphics[width=0.5\linewidth]{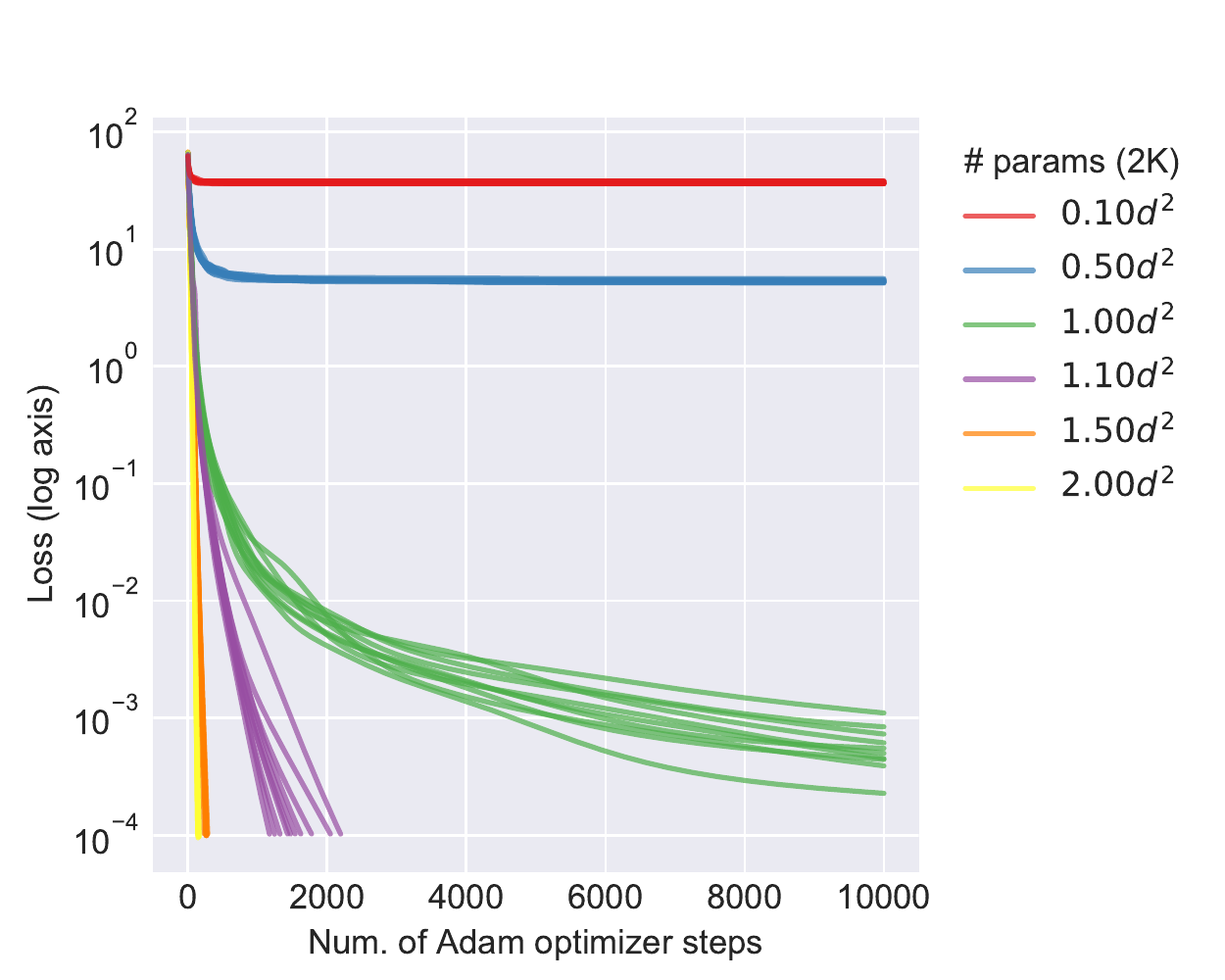}%
}
\caption{a) Experiments using Adam gradient descent for a Haar random target unitary $\mathcal{U}$. b) Experiments using the Adam optimizer for a low-depth target unitary $\mathcal{U}$ of dimension 32 with $8$ parameters ($N$=4).}
\label{fig6}
\end{figure}

\begin{figure}[h]
\subfloat[]{%
  \includegraphics[width=0.5\linewidth]{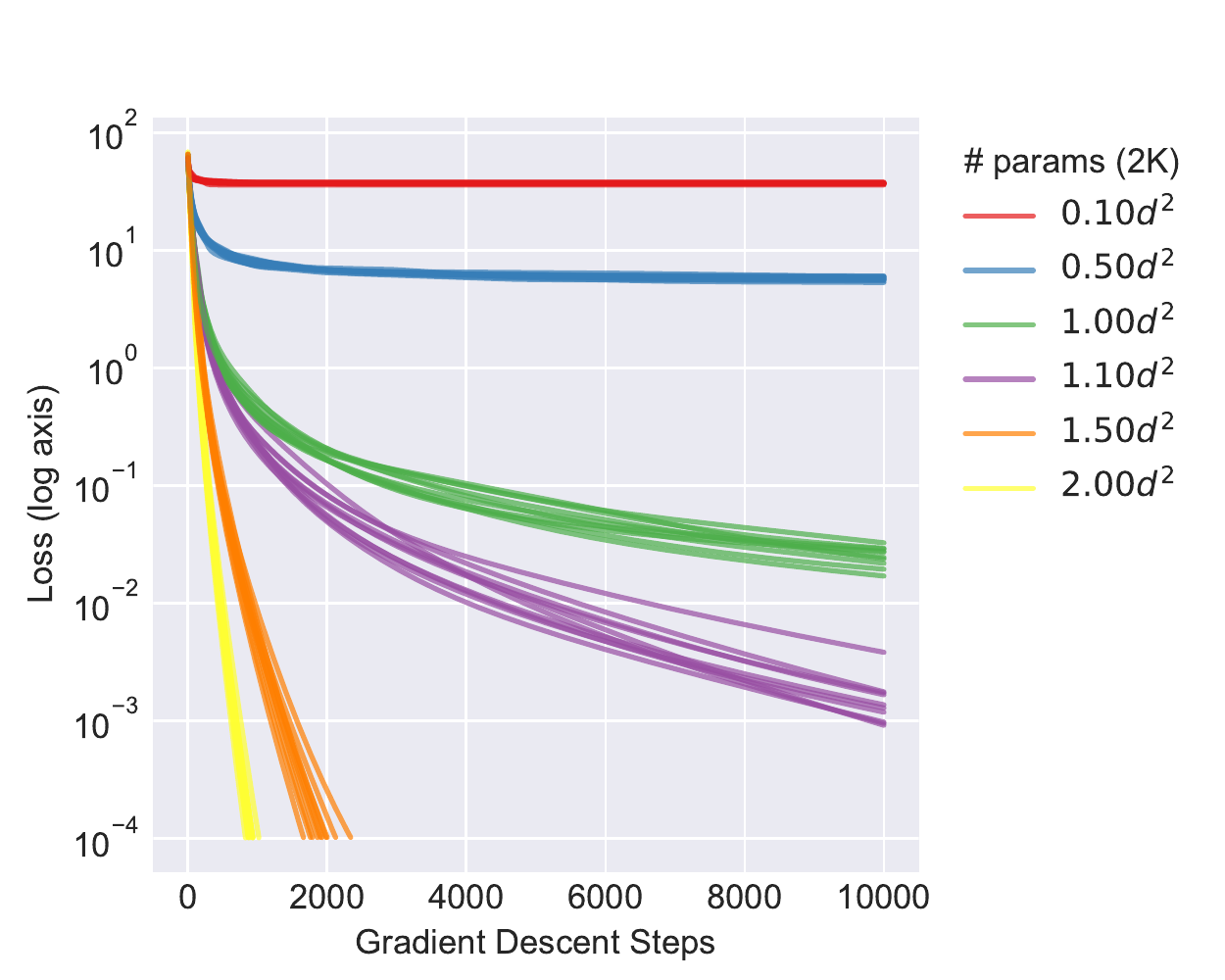}%
}\hfill
\subfloat[]{%
  \includegraphics[width=0.5\linewidth]{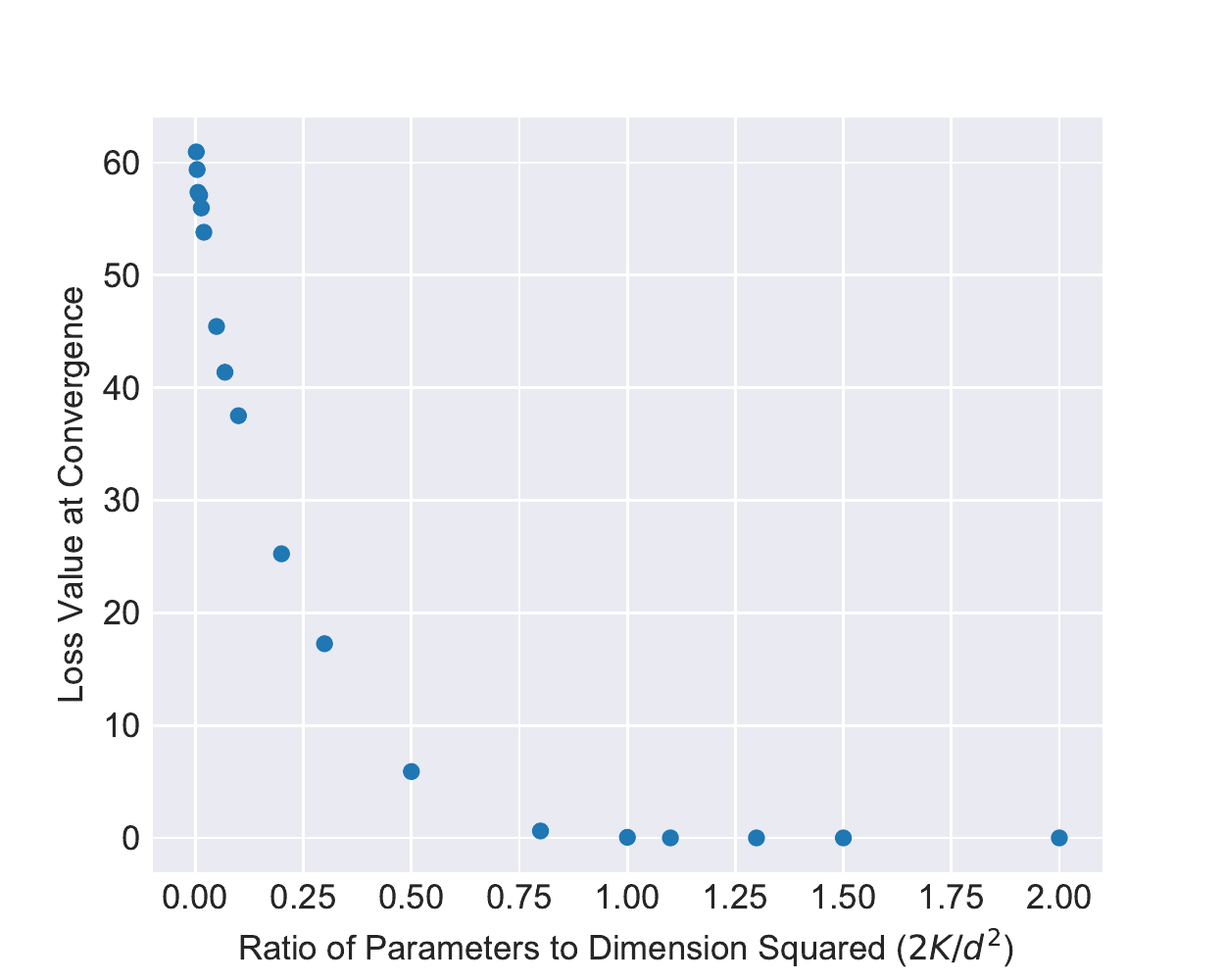}%
}
\caption{a) Simple gradient descent experiments for a target unitary $\mathcal{U}$ of dimension 32 with 8 parameters ($N$=4). b) Value of the loss function after convergence with 10,000 steps of simple gradient descent. Experiments were performed for $\mathcal{U}$ of dimension 32 with various number of parameters.}
\label{fig7}
\end{figure}



In addition to performing optimization using simple (vanilla) gradient descent, we performed optimization using the Adam optimizer \cite{kingma2014adam}, a common optimization method used in deep learning. Adaptive Moment Estimation or Adam is an upgrade of the simple gradient descent algorithm where parameters are assigned different learning rate which are adaptively computed in every iteration of the algorithm. These updates are solely computed from first order gradients. In contrast, the learning rate is fixed for each parameter in simple gradient descent. For more on the Adam optimizer, the reader is referred to \cite{kingma2014adam}. The final loss obtained for learning unitary matrices using the Adam optimizer was consistent with those obtained from simple gradient descent. However, the Adam optimizer appears to converge to a final outcome in far fewer steps. The results of our experiments are provided in Fig.(\ref{fig6}). A comparison between the performance of simple and Adam gradient descent can be observed from Fig.(\ref{fig6}) and Fig.(\ref{fig7}).



\section{Critical points in the under-parameterized models}

\begin{figure}[t]
\includegraphics[width=0.5\linewidth]{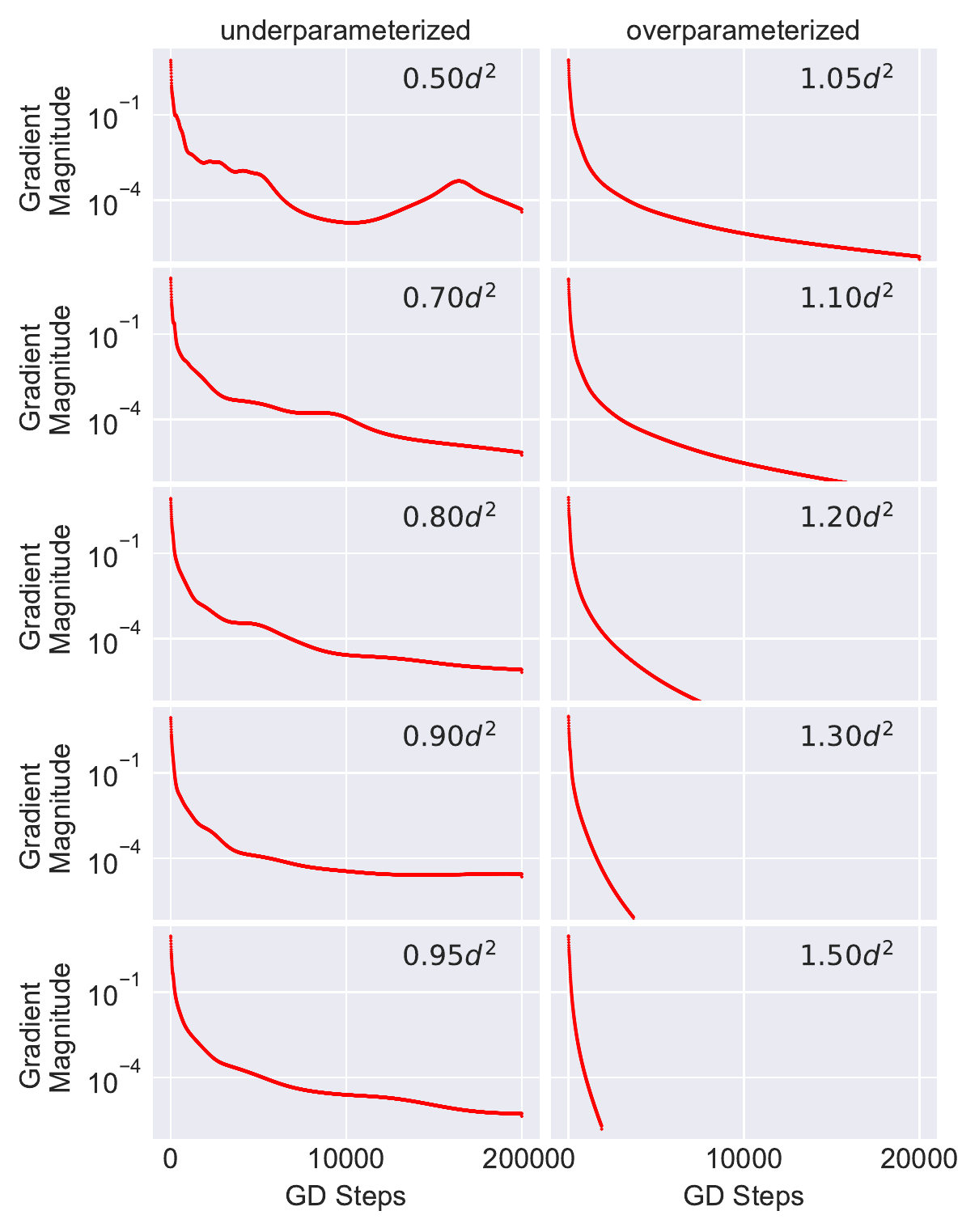}%
\caption{Magnitude of the gradient at each step of gradient descent. In the under-parameterized setting, we find that the magnitudes can increase and decrease over the course of gradient descent. In the over-parameterized setting, we find that the magnitudes decrease, often rapidly, over the course of gradient descent. For each parameter setting, a single experiment was performed which has been plotted here. }
\label{slopes}
 \label{fig8}
 \end{figure}

When learning target unitaries using alternating operator sequences with $d^2$ parameters or more, gradient descent converges to a global minimum of the loss function landscape. When learning with under-parameterized models, we find that gradient descent plateaus at a non-zero loss function value. In the under-parameterized setting, we further explore how the loss function changes over the course of gradient descent by investigating the magnitude of the gradients. In the under-parameterized setting, we find that the magnitude of the gradients can both increase and decrease over the course of gradient descent, suggesting that the path of gradient descent passes in the vicinity of saddle points in the loss landscape. In the over-parameterized setting, the magnitudes of the gradients monotonically decrease with increasing gradient descent steps, suggesting that in this case, the path of gradient descent does not explore saddle points. The results of our findings are presented in Fig.(\ref{fig8}). 

\section{Computation Details}
\label{app:compdetails}
All experiments were performed using the Python package Pytorch \cite{NEURIPS2019_9015}. Experiments were run on a machine equipped with a Nvidia 2080 TI GPU and an Intel Core i7-9700K CPU. Calculations were performed with 64-bit floating point precision. The code used to perform the numerical experiments is available upon request.

\section{A greedy algorithm}
\label{app:greedy}
As noted in the text, we find that gradient descent algorithms require $d^2$ parameters in the sequence $\mathcal{V}(\vec{t},\vec{\tau})$ to learn a low-depth unitary $\mathcal{U}(\vec{t}^*,\vec{\tau}^*) = e^{-i At_N^*} e^{-i B \tau _N^*}  \ldots e^{-i At_1^*} e^{-i B \tau _1^*}$ where $N = O(1)$. This suggests that such low-depth unitaries are intrinsically hard to learn with less than $d^2$ parameters using gradient descent. We also considered a simple greedy algorithm for performing a low-depth target unitary $\mathcal{U}(\vec{t}^*,\vec{\tau}^*)$. Let $\mathcal{V}_q(\vec{t}, \vec{\tau}) = e^{-i At_q} e^{-i B \tau _q}  \ldots e^{-i At_1} e^{-i B \tau _1} $. The first step of the greedy algorithm begins with $q=1$ and uses gradient descent to optimize the parameters $t_1$ and $\tau_1$. The next step of the algorithm at $q=2$ performs gradient descent starting from the initial values $t_2= \tau_2=0$ and $t_1,\tau_1$ which are the optimal values obtained in the previous step. The greedy algorithm then continues, and at each step, $q$ is incremented by 1. At the $q$th step, the initial starting points for gradient descent are $t_q = \tau_q = 0$ and the remaining parameters $\{t_i,\tau_i\}_{1 \leq i \leq q-1}$ are the optimal values obtained at the end of the previous step. We present the pseudocode of the greedy algorithm below.
     
     \begin{algorithm}[H]
     \caption{Greedy Algorithm}
     \label{alg:greedynonconvex}
     
     \textbf{Input:} Low-depth target unitary $\mathcal{U}(\vec{t^*},\vec{\tau^*}) \in U(d)$ and GUE matrices $\pm A,\pm B$.
     
     \vspace{5pt}
     
	 \textbf{Initialize:} Parameters $t_1, \tau_1 = 0$. $\mathcal{V}_0(\vec{t}, \vec{\tau}) = I$. Loss = $a_0= \norm{\mathcal{U}(\vec{t}^*,\vec{\tau}^* ) - I }^2$. 
	 
	 \vspace{3pt}
	 
	 \begin{algorithmic}[1]
	  \While{$q \leq d^2/2$ }
	  
	  \vspace{3pt}
	  
	  \State
	  Construct the unitary $\mathcal{V}_{q-1} e^{-i A t_q} e^{-i B \tau_q}$ with parameters $t_q, \tau_q$. Loss = $a_q= \norm{\mathcal{U}(\vec{t}^*,\vec{\tau}^* ) - \mathcal{V}_{q-1} e^{-i A t_q} e^{-i B \tau_q} }^2$.

	  \vspace{3pt}
	  
	  \State Perform gradient descent on $\{ t_i, \tau_i \}_{1 \leq i \leq q }$ to minimize $a_q$ starting from $\{t'_i, \tau'_i\}_{1 \leq i \leq q-1}$ in $\mathcal{V}_{q-1}$ and $t_q=\tau_q=0$.

      \vspace{3pt}

	  \State Let $\{ t'_i, \tau'_i \}_{1 \leq i \leq q}$ be the optimal parameters. Updated loss = $a'_q= \norm{\mathcal{U}(\vec{t}^*,\vec{\tau}^* ) -  e^{-i A t'_q} e^{-i B \tau'_q} \cdots e^{-i A t'_1} e^{-i B \tau'_1} }^2$.
	  
	  \vspace{3pt}
	  
	  \State \textbf{if} {$a'_q \leq \varepsilon$}, the sequence $\mathcal{V}_q(\vec{t'},\vec{\tau'})$ converges to $\mathcal{U}(\vec{t}^*,\vec{\tau}^* )$ and let Q = q.
	 
	  \vspace{3pt}
     
      \State  \textbf{else} continue.
	  \EndWhile
	 \end{algorithmic}
	  
	 \textbf{Output:}  $\mathcal{V}_Q(\vec{t}',\vec{\tau}')$ such that $\norm{\mathcal{U}(\vec{t}^*,\vec{\tau}^* ) - \mathcal{V}_Q(\vec{t}',\vec{\tau}') }^2 \leq \varepsilon$.
	 
     \end{algorithm}
We investigated the performance of the greedy algorithm for systems of up to five qubits in a restricted area of the loss function landscape. In particular, we considered the parameters $t_j^*,\tau _j^* \in [ -2,2 ]$ in a low-depth target unitary $\mathcal{U}(\vec{t}^*,\vec{\tau}^*)$. In this setting, we find that with a probability that decreases as a function of the number of qubits, the greedy algorithm can construct a sequence $\mathcal{V}_Q(\vec{t},\vec{\tau})$ where $Q \ll d^2/2$. In contrast, gradient descent experiments require $d^2$ parameters in  $\mathcal{V}(\vec{t},\vec{\tau})$ to learn $\mathcal{U}(\vec{t}^*,\vec{\tau}^*)$. That is, the greedy algorithm does indeed sometimes find low-depth target unitaries even in cases where simple gradient descent on under-parameterized sequences fails. For a system of five qubits, the success probability of the greedy algorithm to learn a target unitary $\mathcal{U}(\vec{t}^*,\vec{\tau}^*)$ of depth $N=2$ (\textit{i.e.}, with 4 parameters) using less than 50 parameters in $\mathcal{V}_{Q}(\vec{t},\vec{\tau})$ is around 0.1.

\end{document}